\documentclass[a4paper,fleqn,usenatbib]{mnras}

\usepackage[T1]{fontenc}
\usepackage{ae,aecompl}
\usepackage{graphicx}
\usepackage{amsmath}
\usepackage{amssymb}
\usepackage{subfig}

\newcommand{\Msun}{\ensuremath{\,{M}_\odot}}                            
\newcommand{\Rsun}{\ensuremath{\,{R}_\odot}}                            
\newcommand{\Mjup}{\ensuremath{\,{M}_{\rm Jup}}}                        
\newcommand{\Rjup}{\ensuremath{\,{R}_{\rm Jup}}}                        
\newcommand{\Teq}{\ensuremath{T_{\rm eq}^{\,\prime}}}                   
\newcommand{\safronov}{\ensuremath{\Theta}}                             
\newcommand{\mss}{\,m\,s$^{-2}$}                                        
\newcommand{\FeH}{\ensuremath{\left[\frac{\rm Fe}{\rm H}\right]}}       
\newcommand{\pjup}{\ensuremath{\,\rho_{\rm Jup}}}                       
\newcommand{\psun}{\ensuremath{\,\rho_\odot}}                           

\title[The giant planet WASP-36\,b]{An optical transmission spectrum of the giant planet WASP-36\,b}

\author[L. Mancini et al.]{L. Mancini$^{1,\,2}$\thanks{E-mail: mancini@mpia.de},
J. Kemmer$^{1}$,
J. Southworth$^{3}$,
K. Bott$^{4}$,
P. Molli\`{e}re$^{1}$,
S. Ciceri$^{1}$,
\newauthor
G. Chen$^{5}$,
Th. Henning$^{1}$\\
\\
$^{1}$Max Planck Institute for Astronomy, K\"{o}nigstuhl 17, 69117 -- Heidelberg, Germany \\
$^{2}$INAF -- Osservatorio Astrofisico di Torino, via Osservatorio 20, 10025 -- Pino Torinese, Italy \\
$^{3}$Astrophysics Group, Keele University, Keele ST5 5BG, UK \\
$^{4}$Exoplanetary Science at UNSW, Australian Centre for Astrobiology, School of Physics, UNSW Australia \\
$^{5}$Key Laboratory of Planetary Sciences, Purple Mountain Observatory, Chinese Academy of Sciences, Nanjing 210008, China \\
}

\date{Accepted XXX. Received YYY; in original form ZZZ}

\pubyear{2015}

\begin{document}
\label{firstpage}
\pagerange{\pageref{firstpage}--\pageref{lastpage}}
\maketitle

\begin{abstract}
We present broad-band photometry of five transits in the planetary system WASP-36, totaling 17 high-precision light curves. Four of the transits were simultaneously observed in four passbands ($g^{\prime}, r^{\prime}, i^{\prime}, z^{\prime}$), using the telescope-defocussing technique, and achieving scatters of less than $1$\,mmag per observation. We used these data to improve the measured orbital and physical properties of the system, and obtain an optical transmission spectrum of the planet. We measured a decreasing radius from bluer to redder passbands with a confidence level of more than $5\sigma$. The radius variation is roughly 11 pressure scale heights between the $g^{\prime}$ and the $z^{\prime}$ bands. This is too strong to be Rayleigh scattering in the planetary atmosphere, and implies the presence of a species which absorbs strongly at bluer wavelengths.
\end{abstract}



\section{Introduction}
\label{sect_01}

Transiting hot Jupiters are a class of exoplanets which are very suitable for detailed study of their physical and orbital parameters. They are gas-giant planets with short orbital periods (mass $M_{\rm p}>0.3 M_{\rm Jup}$ and period $P < 10$\,d), and many of them are now known orbiting bright stars. Their relatively large sizes usually give deep transits (typically 0.5--3.5\%) which are well-suited for observation and analysis. Another particular advantage of transiting hot Jupiters is their suitability for \emph{transmission spectroscopy}, which can yield constraints on the chemical composition at the terminator of their atmospheres from transit depth measurements at multiple wavelengths. Depending on the temperature of the planet's atmosphere, it is possible to investigate the presence of several molecules (e.g.\ H$_2$O, CO, CO$_2$ and CH$_4$) at infrared (IR) wavelengths. The blue region of the visual wavelength range is also important to study, as it allows constraints to be placed on the presence of Rayleigh scattering or particular atomic absorption features (e.g.\ Na).
There has been significant progress in this field in the last few years, and many exoplanets have been observed using transmission spectroscopy. The resulting transmission spectra show an unexpectedly diversity of the atmospheres of hot Jupiters, in particular the amount of cloud at observable pressure levels \citep{sing:2016}.

Whilst most of these studies were performed at IR wavelengths, a small set of planets have also been studied at optical wavelengths from space with the Hubble and the Spitzer telescopes (i.e.\ HD\,189733\,b: \citealp{pont:2013}; HD\,209458\,b: \citealp{desert:2008}; HAT-P-1\,b: \citealp{nikolov:2014}; HAT-P-12\,b: \citealp{sing:2016}; WASP-6\,b: \citealp{nikolov:2015}; WASP-12\,b: \citealp{sing:2013}; WASP-17\,b: \citealp{sing:2016}; WASP-19\,b: \citealp{huitson:2013}; WASP-31\,b: \citealp{sing:2015}; WASP-39\,b: \citealp{fischer:2016};) and from large ground-based telescopes such as the Very Large Telescope (GJ\,1214\,b: \citealp{bean:2010,bean:2011}; WASP-19\,b: \citealp{sedaghati:2015}), the Gran Telescopio CANARIAS (HAT-P-19\,b: \citealp{mallonn:2015}; WASP-43\,b: \citealp{murgas:2014}), Gemini (HAT-P-32: \citealp{gibson:2013a}; WASP-12: \citealp{stevenson:2014}; WASP-29: \citealp{gibson:2013b}) and Magellan (HAT-P-26: \citealp{stevenson:2015}; TrES-3: \citealp{parviainen:2016}; WASP-6: \citealp{jordan:2013}).

An alternative approach for probing planetary atmospheres is that of \emph{transmission photometry}, which has a lower spectral resolution but is suitable for ground-based telescopes with smaller apertures and exoplanets orbiting faint stars. Moreover, photometric observations are much less affected by telluric contamination than spectroscopic ones.

In this context, we are carrying out a large program of transit observations of the known hot Jupiters with an array of medium-size telescopes in both hemispheres \citep{mancini:2016}. In particular, we are using GROND (Gamma-Ray Burst Optical and Near-Infrared Detector), an imaging camera able obtain light curves in four optical and three near-IR passbands simultaneously. These are used to improve measurements of the physical properties of the planets and their host stars, and to look for transit-depth variations as a function of wavelength in the optical wavelength region.

In order to have a comprehensive picture, we are investigating both inflated and compact hot Jupiters. Due to their density, the latter should be not optimal targets for transmission-spectrum studies. However, depending on the equilibrium temperature, opacity and chemical composition of their atmospheres, Mie scattering, Rayleigh scattering and molecular opacity could still cause strong variation of the radius of such planets, especially if we compare measurements at the wavelength region $400-500$\,nm with those at $800-900$\,nm.

Previous studies with GROND have shown different atmospheric properties: we obtained transmission photometry consistent with flat spectra for several systems (WASP-23\,b: \citealp{nikolov:2013}; WASP-43\,b: \citealp{chen:2014}; WASP-80\,b: \citealp{mancini:2014a}; WASP-67\,b: \citealp{mancini:2014b}), and larger planetary radii at bluer wavelengths for two objects (Qatar-2\,b: \citealp{mancini:2014c}; WASP-103: \citealp{southworth:2015}). These two latter studies were based on multiple datasets from GROND and demonstrate that repeated transit observations produce more precise and robust results. It is interesting to note the large variation that was found in the dense and moderately cold planet Qatar-2\,b.

In the present work, we study the transiting planetary system WASP-36 \citep{smith:2012}. This contains a relatively dense hot Jupiter, WASP-36\,b, with mass $M_{\rm p}\approx 2.3\,M_{\rm Jup}$ and radius $R_{\rm p}\approx 1.3\,R_{\rm Jup}$. The density of WASP-36\,b is therefore similar to that of Qatar-2\,b, but its temperature is higher since it orbits a hotter ($5959 \pm 134$\,K) and metal-poor ([Fe/H]$=-0.26\pm0.10$) G2\,V star, WASP-36\,A, every 1.54\,days. The high temperature and low densities imply that the planet atmosphere would be rich of molecules consisting of hydrogen and sulfur, which are able to absorb stellar light at blue-optical wavelengths.

An occultation event of WASP-36\,b was observed in the $K_{\rm s}$ band by \citet{zhou:2015}. These authors found that the orbit is consistent with circular and that the occultation is $0.13 \pm 0.04$\% deep, corresponding to a $K_{\rm s}$-band brightness temperature of $T_{\rm B} = 1900^{+100}_{-200}$\,K. This is similar to the planet's equilibrium temperature ($T^{\prime}_{\rm eq}\approx1725$\,K) and means that strong absorbers are expected to be present in its atmosphere \citep{fortney:2010}. \citet{maciejewski:2016} has recently presented a new photometric light curve of a WASP-36\,b transit, which was used to redetermine the parametrs of the planetary system.

The paper is structured as follows. The observations and data reduction are both described in Sect.\,\ref{sec_2}, while the analysis of the data is presented in Sect.\,\ref{sec_3}. The refinement of the orbital ephemerides is given in Sect.\,\ref{sec_3}. In Sect.\,\ref{sec_4} we revise the main physical properties of the planetary system. In Sect.\,\ref{sec_5} we investigate the variation of the planetary radius as function of wavelength and, finally, we summarize our results in Sect.\,\ref{sec_6}.


\begin{figure}
\centering
\includegraphics[width=\columnwidth]{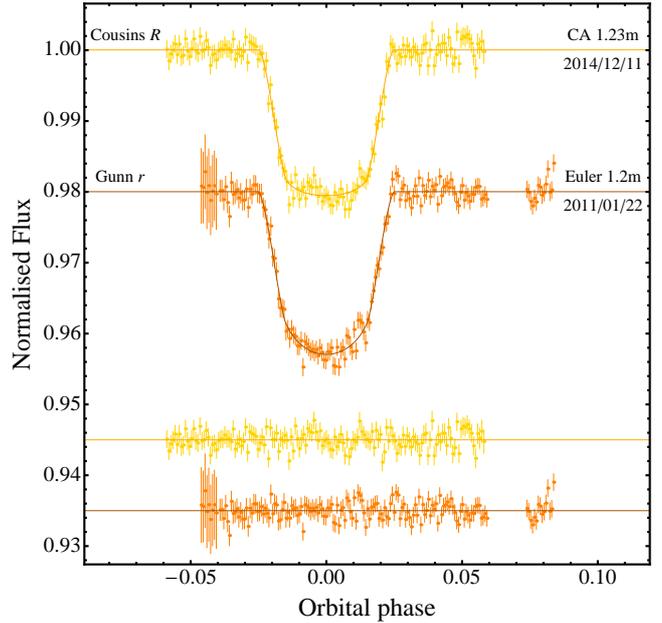}
\caption{Transit light curve from Calar Alto (upper; this work) and the best one from the discovery paper (lower; \citealt{smith:2012}).
The dates and filters used are indicated. The solid lines show the {\sc jktebop} best fits and the residuals are shown at the bottom.}
\label{fig:CA_lc}
\end{figure}
\begin{figure}
\centering
\includegraphics[width=\columnwidth]{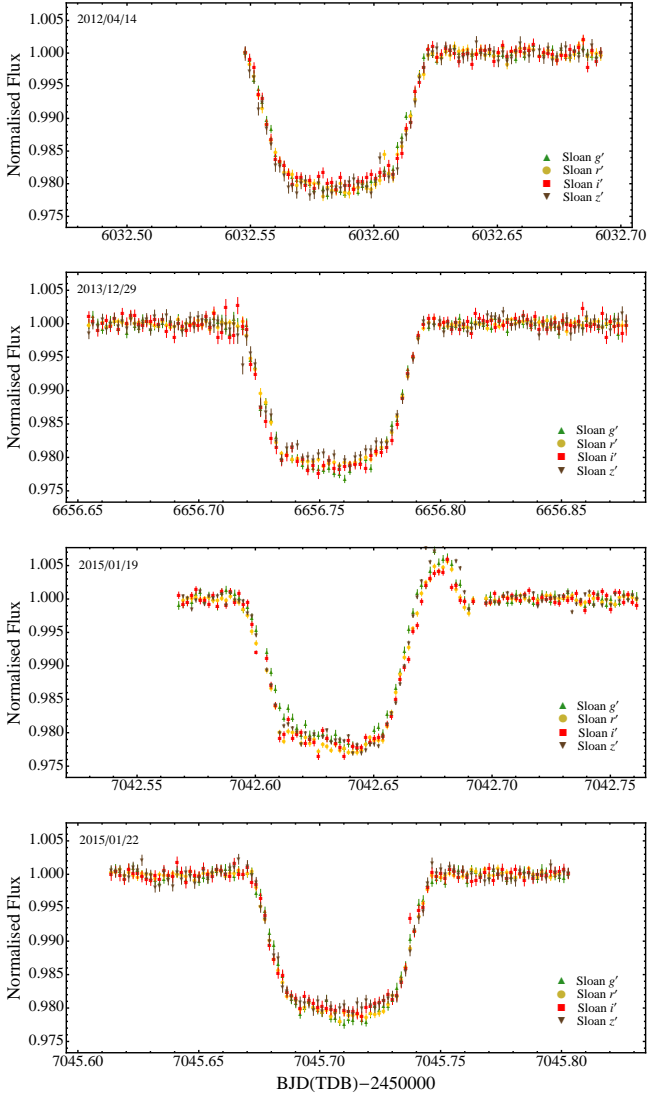}
\caption{The four transit light curves from GROND, ordered according to date.}
\label{fig:GROND_lc_1}
\end{figure}
\begin{figure}
\centering
\includegraphics[width=\columnwidth]{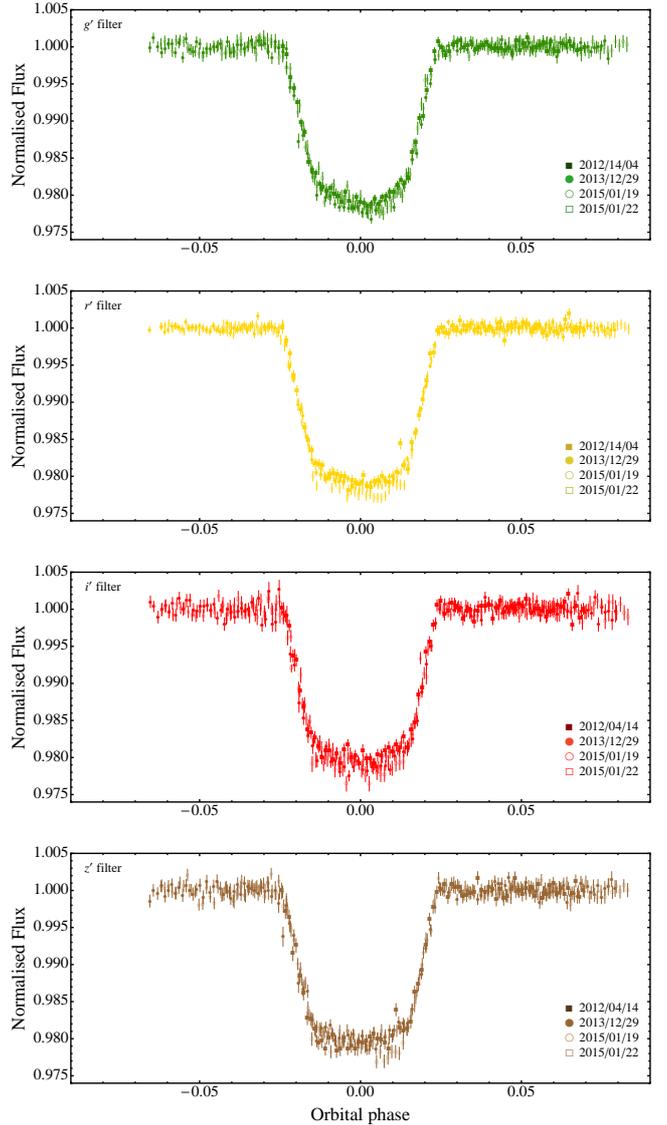}
\caption{As Fig.\,\ref{fig:GROND_lc_1} but sorted according to filter.}
\label{fig:GROND_lc_2}
\end{figure}
\begin{figure}
\centering
\includegraphics[width=\columnwidth]{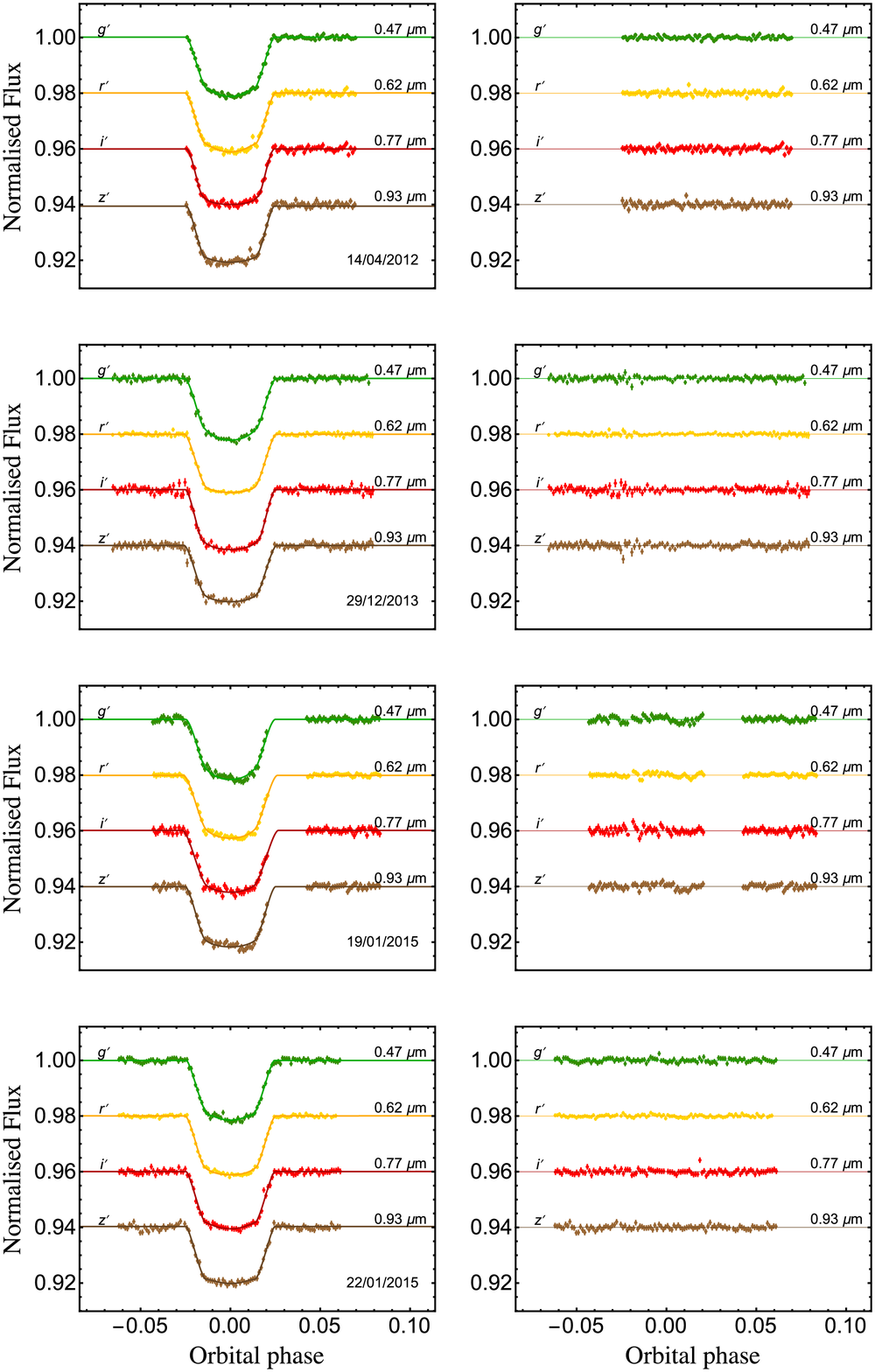}
\caption{Phased light curves of WASP-36 from GROND compared to the {\sc jktebop}
best fits. The passbands and central wavelengths are labelled.}
\label{fig:GROND_lc_3}
\end{figure}

\section{Observation and data reduction}
\label{sec_2}

Four transits by WASP-36\,b were observed with GROND, which is mounted on the MPG\,2.2\,m telescope at ESO La Silla (Table\,\ref{tab:obs}). GROND is a seven-channel imaging camera that was built for rapid observations of gamma-ray burst afterglows \citep{greiner:2008}, but is also well-suited for transit observations. GROND can be used to observe simultaneously in four optical passbands (similar to Sloan $g^{\prime}, r^{\prime}, i^{\prime}, z^{\prime}$) and three near-IR bands ($J$, $H$, $K$). The optical light is collected by back-illuminated $2048 \times 2048$ pixel E2V CCDs, with a field of view of 5.4\,arcmin $\times$ 5.4\,arcmin at a scale of 0.158\,arcsec\,pixel$^{-1}$. The near-IR channels use $1024 \times 1024$ pixel Rockwell HAWAII-1 arrays, with a field of view of 10\,arcmin $\times$ 10\,arcmin at 0.6\,arcsec\,pixel$^{-1}$. The photometric precision of the near-IR arms is significantly worse than the optical ones \citep{pierini:2012}, so these data will be not considered in this work.

Another transit by WASP-36\,b was remotely observed in December 2014 through a Cousins $R$ filter with the Zeiss 1.23\,m telescope at the Calar Alto Observatory. The telescope is equipped with a $4096\times 4096$ pixel DLR-MKIII camera, with a field of view of 21.5\,arcmin $\times$ 21.5\,arcmin at a plate scale of 0.32\,arcsec\,pixel$^{-1}$. The CCD was operated without binning, but windowed to shorten the readout time. During all the observations the two telescopes were autoguided and defocussed in order to improve the photometric precision of the data.

\begin{table*}
\centering
\setlength{\tabcolsep}{4pt}
\caption{Details of the transit observations presented in this work. $N_{\rm obs}$ is the number of observations,
$T_{\rm exp}$ is the exposure time, $T_{\rm obs}$ is the observational cadence, and `Moon illum.' is the geocentric
fractional illumination of the Moon at midnight (UT). The aperture sizes are the radii of the software apertures
for the star, inner sky and outer sky, respectively. Scatter is the \emph{rms} scatter of the data versus a fitted model.}
\label{tab:obs}
\begin{tabular}{lccccccccccc} \hline
Telescope & Date of   & Start time & End time  &$N_{\rm obs}$ & $T_{\rm exp}$ & $T_{\rm obs}$ & Filter & Airmass & Moon & Aperture & Scatter \\
               & first obs &    (UT)    &   (UT)    &              & (s)           & (s)           &        &         &illum.& radii (px) & (mmag)  \\
\hline
MPG\,2.2\,m & 2012 04 15 & 01:44 & 04:36 &  82 &  90 & 156 & Sloan $g^{\prime}$ & $1.12 \rightarrow 2.70$                  &  23\%  & 30,53,90  & 0.54 \\
MPG\,2.2\,m & 2012 04 15 & 01:44 & 04:36 &  82 &  90 & 156 & Sloan $r^{\prime}$ & $1.12 \rightarrow 2.70$                  &  23\%  & 32,68,85  & 0.69 \\
MPG\,2.2\,m & 2012 04 15 & 01:44 & 04:36 &  82 &  90 & 156 & Sloan $i^{\prime}$ & $1.12 \rightarrow 2.70$                  &  23\%  & 25,55,80  & 0.76 \\
MPG\,2.2\,m & 2012 04 15 & 01:44 & 04:36 &  82 &  90 & 156 & Sloan $z^{\prime}$ & $1.12 \rightarrow 2.70$                  &  23\%  & 30,58,90  & 0.84 \\[2pt]
MPG\,2.2\,m & 2013 12 30 & 03:42 & 08:57 & 121 & 100 & 142 & Sloan $g^{\prime}$ & $1.85 \rightarrow 1.07 \rightarrow 1.23$ & ~~2\%  & 27,65,90  & 0.69 \\
MPG\,2.2\,m & 2013 12 30 & 03:42 & 08:57 & 122 & 100 & 142 & Sloan $r^{\prime}$ & $1.85 \rightarrow 1.07 \rightarrow 1.23$ & ~~2\%  & 27,65,90  & 0.44 \\
MPG\,2.2\,m & 2013 12 30 & 03:42 & 08:57 & 124 & 100 & 142 & Sloan $i^{\prime}$ & $1.85 \rightarrow 1.07 \rightarrow 1.23$ & ~~2\%  & 19,61,90  & 0.88 \\
MPG\,2.2\,m & 2013 12 30 & 03:42 & 08:57 & 121 & 100 & 142 & Sloan $z^{\prime}$ & $1.85 \rightarrow 1.07 \rightarrow 1.23$ & ~~2\%  & 32,63,91  & 0.87 \\[2pt]
CA\,1.23\,m & 2014 12 11 & 01:44 & 06:02 & 154 &  90 & 100 & Cousins $R$        & $1.66 \rightarrow 1.42 \rightarrow 1.79$ &  80\%  & 23,33,60  & 1.21 \\[2pt]
MPG\,2.2\,m & 2015 01 20 & 01:37 & 06:16 &  88 & 110 & 156 & Sloan $g^{\prime}$ & $2.04 \rightarrow 1.07 \rightarrow 1.08$ & ~~0\%  & 26,66,100 & 0.83 \\
MPG\,2.2\,m & 2015 01 20 & 01:37 & 06:16 &  88 & 110 & 156 & Sloan $r^{\prime}$ & $2.04 \rightarrow 1.07 \rightarrow 1.08$ & ~~0\%  & 24,62,100 & 0.61 \\
MPG\,2.2\,m & 2015 01 20 & 01:37 & 06:16 &  87 & 110 & 156 & Sloan $i^{\prime}$ & $2.04 \rightarrow 1.07 \rightarrow 1.08$ & ~~0\%  & 24,60,90  & 0.95 \\
MPG\,2.2\,m & 2015 01 20 & 01:37 & 06:16 &  87 & 110 & 156 & Sloan $z^{\prime}$ & $2.04 \rightarrow 1.07 \rightarrow 1.08$ & ~~0\%  & 29,66,100 & 0.83 \\[2pt]
MPG\,2.2\,m & 2015 01 23 & 02:43 & 07:16 & 105 & 110 & 156 & Sloan $g^{\prime}$ & $1.39 \rightarrow 1.07 \rightarrow 1.19$ & ~~10\% & 34,94,110 & 0.65 \\
MPG\,2.2\,m & 2015 01 23 & 02:43 & 07:16 & 101 & 110 & 156 & Sloan $r^{\prime}$ & $1.39 \rightarrow 1.07 \rightarrow 1.19$ & ~~10\% & 29,55,97  & 0.43 \\
MPG\,2.2\,m & 2015 01 23 & 02:43 & 07:16 & 104 & 110 & 156 & Sloan $i^{\prime}$ & $1.39 \rightarrow 1.07 \rightarrow 1.19$ & ~~10\% & 30,53,88  & 0.76 \\
MPG\,2.2\,m & 2015 01 23 & 02:43 & 07:16 & 103 & 110 & 156 & Sloan $z^{\prime}$ & $1.39 \rightarrow 1.07 \rightarrow 1.19$ & ~~10\% & 31,59,107 & 0.75 \\
\hline \end{tabular} \end{table*}

The reduction of the data was performed using the {\sc defot} pipeline, written in {\sc idl}\footnote{The acronym IDL stands for Interactive Data Language and is a trademark of ITT Visual Information Solutions.} \citep{southworth:2014}. Briefly, the scientific images were calibrated using master bias and flat-field frames, produced by median-combining individual calibration images. The target and a suitable set of non-variable comparison stars were identified in a reference image, which was used to measure pointing variations by a cross-correlation process. Three apertures were placed by hand around the selected stars and the aperture radii were chosen to obtain the lowest scatter versus a fitted model. Differential photometry was obtained using the {\sc aper} routine\footnote{{\sc aper} is part of the {\sc astrolib} subroutine library distributed by NASA.} and rectified to zero magnitude by fitting a straight line to the out-of-transit data. The data will be made available at the CDS\footnote{{\tt http://cdsweb.u-strasbg.fr/}}.

The Calar Alto light curve is plotted in Fig.\,\ref{fig:CA_lc} together with the best one reported in the discovery paper \citep{smith:2012} for comparison. The GROND light curves are plotted according to date in Fig.\,\ref{fig:GROND_lc_1} and to filter in Fig.\,\ref{fig:GROND_lc_2}. A clear anomaly is visible in the GROND data observed on January 19, 2015 (third panel in Fig.\,\ref{fig:GROND_lc_1}). This variation is not correlated with the position of the stars on the CCDs, the CCD temperatures, the weather conditions, or vignetting from the dome slit. There is also no obvious astrophysical explanation. We therefore assumed that the anomaly was caused by an unknown instrumental effect and removed the affected data from our analysis.

\section{Light curve analysis}
\label{sec_3}

We have modelled our new light curves and the best one from \citet{smith:2012}, which was observed with the Euler 1.2\,m telescope, using the {\sc jktebop}\footnote{\textsc{jktebop} is written in {\sc fortran77} and is available at: {\tt http://www.astro.keele.ac.uk/jkt/codes/jktebop.html}} code (\citealp{southworth:2013}). Each light curve was fitted using the following parameters: the sum and ratio of the fractional radii\footnote{The fractional radii are defined as $r_{\mathrm{A}} = R_{\mathrm{A}}/a$ and $r_{\mathrm{b}} = R_{\mathrm{b}}/a$, where $R_{\mathrm{A}}$ and $R_{\mathrm{b}}$ are the true radii of the star and planet, and $a$ is the semi-major axis.} ($r_{\mathrm{A}}+r_{\mathrm{b}}$ and $k=r_{\mathrm{b}}/r_{\mathrm{A}}$), the orbital period and inclination ($P$ and $i$), the time of transit midpoint ($T_{0}$) and the linear coefficient of the quadratic limb darkening law ($u_{\mathrm{A}}$). The second limb darkening coefficient ($v_{\mathrm{A}}$) was fixed to a theoretical value \citep{claret:2004}. We assumed that the orbit of WASP-36\,b is circular \citep{smith:2012,zhou:2015}.

\begin{figure}
\centering
\includegraphics[width=8.0cm]{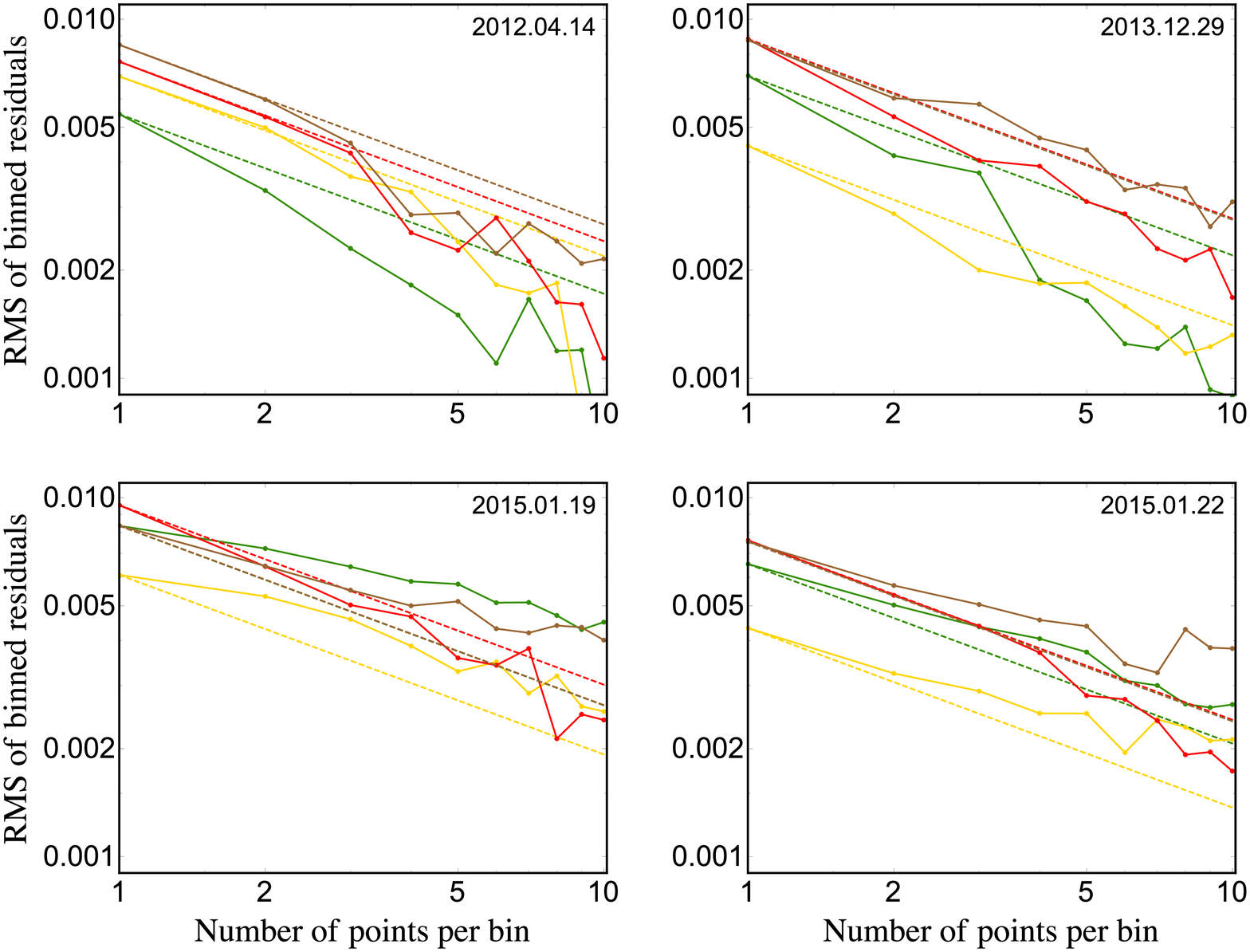}
\caption{Root mean square of binned residuals versus the number of points per bin for each of the four transits of WASP-36 observed with GROND and for each filter: green line is $g^{\prime}$, yellow line is $r^{\prime}$, red line is $i^{\prime}$ and brown line is $z^{\prime}$. The dashed lines are proportional to $N^{-1/2}$ and are normalized to match the value for bin size $N = 1$ for each light curve.}
\label{fig:precisions}
\end{figure}

The {\sc aper} algorithm, used in our reduction pipeline, usually underestimates the uncertainties in the differential magnitudes. Red noise also affects time-series photometry as it is not accounted for by standard error estimation algorithms (e.g.\ \citealp{carter:2009}). We therefore rescaled the error bars to give a reduced $\chi^{2}$ of $\chi_{\nu}^{2}= 1$ (e.g.\ \citealp{mancini:2014a,southworth:2015}).  Moreover, using the likelihood function defined as in Eq.\,(32) of \citet{carter:2009}, we performed a wavelet-basis red-noise MCMC analysis for estimating the Gaussian white noise and the correlated red noise for each light curve. They are reported in the last two columns of Table\,\ref{tab:photometry}. Fig.\,\ref{fig:precisions} shows the root mean square of the binned residuals versus the bin size, which is another way to illustrate the photometric precision that we achieved.

\subsection{Orbital period determination}
\label{sec_3.1}

\begin{figure*} \centering
\includegraphics[width=\textwidth]{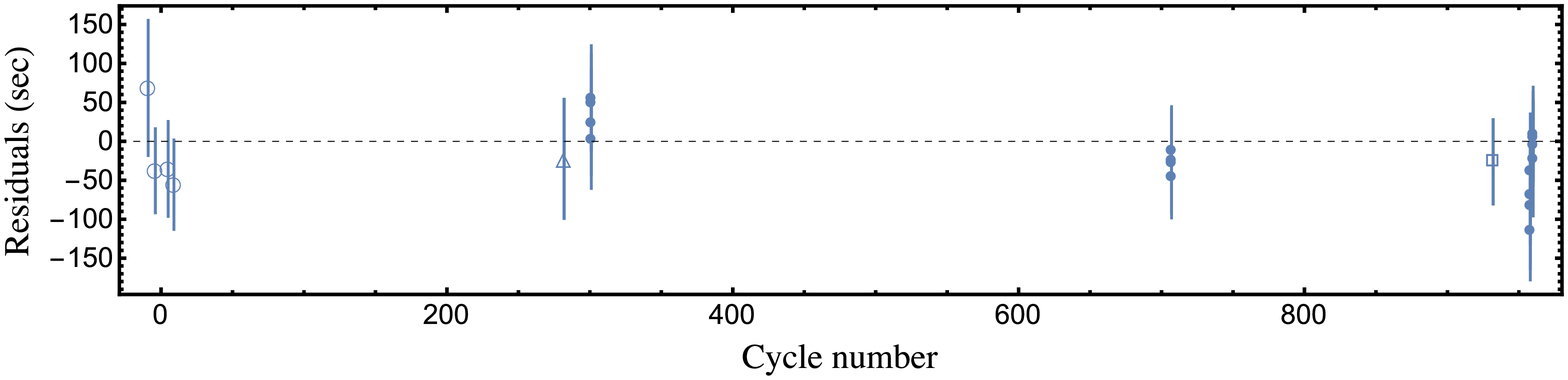}
\caption{Residuals of the times of mid-transit versus a linear ephemeris. The timings from \citet{smith:2012} 
are plotted using open circles, from \citet{maciejewski:2016}  with a triangle, from GROND with filled circles, and from Calar Alto with an open box.}
\label{fig:O-C}
\end{figure*}

We estimated the times of mid-transit by fitting each light curve with {\sc jktebop}. The uncertainties were obtained using Monte Carlo simulations. 
We also considered the timing measured by \citet{smith:2012} and \citet{maciejewski:2016}.
All transit timings (see Table\,\ref{tab:tmin}) were then fitted with a straight line to obtain the following orbital ephemeris:
\begin{equation}
T_{0} = \,$BJD(TDB)$\,2455569.83771(46) + 1.53736596(24)\,E , \nonumber
\end{equation}
where $E$ denotes the number of orbital cycles after the reference epoch, and the quantities in brackets denote the uncertainty in the final digit of the preceding number. The residuals of the fit are plotted in Fig.\,\ref{fig:O-C}. The linear ephemeris is not a good match to the observations -- the fit has $\chi_{\nu}^{2}= 3.76$ -- so the uncertainties have been increased to account for this. As our timings comprise only ten epochs over more than 1000 orbits, and in line with previous work (e.g.\ \citealp{southworth:2014,ciceri:2015,mancini:2015}), we do not interpret this as an indication of transit time variations.
\begin{table}
\setlength{\tabcolsep}{4pt}
\centering
\caption{Times of transit midpoint of WASP-36 and their residuals.}
\label{tab:tmin}
\begin{tabular}{lrrc}
\hline
~~~~Time of minimum  & Cycle & O-C~~~  & Reference  \\
~~BJD(TDB)$-2400000$ & no.~  & (JD)~~~ &            \\
\hline \\[-8pt]%
$5556.00221 \pm         0.00055$ &           -9 &          0.00075091 & \citet{smith:2012} \\
$5563.68781 \pm         0.00017$ &           -4 &         -0.00047861 & \citet{smith:2012} \\
$5577.52413 \pm         0.00025$ &            5 &         -0.00045174 & \citet{smith:2012} \\
$5583.67336 \pm         0.00021$ &            9 &         -0.00068535 & \citet{smith:2012} \\ [2.pt]
$6003.37465 \pm         0.00043$ &        282 &          0.00013033 &  \citet{maciejewski:2016}  \\ [2.pt]
$6032.58490 \pm         0.00028$ &        301 &          0.00001463 & $g^{\prime}$ MPG\,2.2\,m        \\
$6032.58515 \pm         0.00031$ &        301 &          0.00025994   & $r^{\prime}$ MPG\,2.2\,m        \\
$6032.58545 \pm         0.00025$ &        301 &          0.00055714   & $i^{\prime}$ MPG\,2.2\,m        \\
$6032.58551 \pm         0.00031$ &        301 &          0.00062418   & $z^{\prime}$ MPG\,2.2\,m        \\ [2.pt]
$6656.75514 \pm         0.00024$ &        707 &         -0.00030923    & $g^{\prime}$ MPG\,2.2\,m        \\
$6656.75532 \pm         0.00012$ &        707 &         -0.00012608   & $r^{\prime}$ MPG\,2.2\,m        \\
$6656.75493 \pm         0.00014$ &        707 &         -0.00051748   & $i^{\prime}$ MPG\,2.2\,m        \\
$6656.75517 \pm         0.00030$ &        707 &         -0.00027314   & $z^{\prime}$ MPG\,2.2\,m        \\ [2.pt]
$7002.66248 \pm         0.00012$ &        932 &         -0.00029382   & \,\,\, CA\,1.23\,m        \\ [2.pt]
$7042.63387 \pm         0.00033$ &        958 &         -0.00041580   & $g^{\prime}$ MPG\,2.2\,m        \\
$7042.63299 \pm         0.00024$ &        958 &         -0.00130137    & $r^{\prime}$ MPG\,2.2\,m        \\
$7042.63336 \pm         0.00044$ &        958 &         -0.00093176   & $i^{\prime}$ MPG\,2.2\,m        \\
$7042.63352 \pm         0.00022$ &        958 &         -0.00076909   & $z^{\prime}$ MPG\,2.2\,m        \\ [2.pt]
$7045.70910 \pm         0.00015$ &        960 &          0.00008466   & $g^{\prime}$ MPG\,2.2\,m        \\
$7045.70899 \pm         0.00010$ &        960 &         -0.00002979  & $r^{\prime}$ MPG\,2.2\,m        \\
$7045.70878 \pm         0.00035$ &        960 &         -0.00023732   & $i^{\prime}$ MPG\,2.2\,m        \\
$7045.70915 \pm         0.00018$ &        960 &          0.00013033   & $z^{\prime}$ MPG\,2.2\,m        \\
\hline \end{tabular} \end{table}

\subsection{Photometric parameters}
\label{sec_3.2}

The best fits to the light curves are shown in Fig.\,\ref{fig:CA_lc} and Fig.\,\ref{fig:GROND_lc_3}. The parameters of the fits are reported in Table\,\ref{tab:photometry}. The uncertainties of the parameters were estimated for each solution from 10\,000 Monte Carlo simulations and through a residual-permutation algorithm. The larger of the two error bars was adopted in each case. The final photometric parameters were calculated as the weighted mean of the results in Table\,\ref{tab:photometry}. We also show the values obtained by \citet{smith:2012} and \citet{maciejewski:2016} for comparison.
\begin{table*} 
\centering
{\tiny
\caption{Photometric properties of the WASP-36 system derived by fitting the light curves. $\sigma_{\rm white}$ is the Gaussian white noise, while $\sigma_{\rm red}$ indicates the correlated red noise, which is defined in \citet{carter:2009}.}
\label{tab:photometry}
\begin{tabular}{llllcllcc}
\hline
Source & ~Filter & $~~~~r_{\mathrm{A}}+r_{\mathrm{b}}$& $~~~~~~~k$ & $i^{\circ}$ & $~~~~~~r_{\mathrm{A}}$ &$~~~~~~r_{\mathrm{b}}$ & $\sigma_{\rm red}$ ($10^{-3}$) & $\sigma_{\rm white}$ ($10^{-3}$) \\
\hline
MPG\,2.2\,m/GROND\,\#1 & Sloan $g^{\prime}$ & $0.1908 \pm 0.0047$ & $0.1411 \pm 0.0051$ & $83.34 \pm 0.34$ & $0.1672 \pm 0.0041$ & $0.02359 \pm 0.00091$ & 0.97 $^{+  1.07}_{-  0.66}$ &   0.41 $^{+  0.05}_{-  0.06}$ \\[4pt]
MPG\,2.2\,m/GROND\,\#1 & Sloan $r^{\prime}$ & $0.1917 \pm 0.0087$ & $0.1387 \pm 0.0077$ & $83.47 \pm 0.66$ & $0.1683 \pm 0.0076$ & $0.0233~  \pm 0.0015$  & 1.17 $^{+  1.20}_{-  0.79}$ &   0.57 $^{+  0.07}_{-  0.07}$ \\[4pt]
MPG\,2.2\,m/GROND\,\#1 & Sloan $i^{\prime}$ & $0.178 ~ \pm 0.012$  & $0.128 ~ \pm 0.010$  & $84.64 \pm 0.95$ & $0.158~  \pm 0.011$  & $0.0203~  \pm 0.0017$  & 1.86 $^{+  1.48}_{-  1.19}$ &   0.49 $^{+  0.07}_{-  0.11}$\\[4pt]
MPG\,2.2\,m/GROND\,\#1 & Sloan $z^{\prime}$ & $0.189 ~ \pm 0.012$  & $0.135 ~ \pm 0.014$  & $83.50 \pm 0.77$ & $0.167~  \pm 0.012$  & $0.0225~  \pm 0.0015$  & 1.58 $^{+  1.30}_{-  1.00}$ &   0.63 $^{+  0.08}_{-  0.08}$\\[6pt]
MPG\,2.2\,m/GROND\,\#2 & Sloan $g^{\prime}$ & $0.194 ~ \pm 0.010$  & $0.1363 \pm 0.0033$ & $83.62 \pm 0.89$ & $0.1704 \pm 0.0084$ & $0.0232~  \pm 0.0017$  & 1.15 $^{+  1.11}_{-  0.79}$ &   0.64 $^{+  0.07}_{-  0.06}$\\[4pt]
MPG\,2.2\,m/GROND\,\#2 & Sloan $r^{\prime}$ & $0.204 ~ \pm 0.013$  & $0.136 ~ \pm 0.010$  & $82.56 \pm 0.49$ & $0.180~  \pm 0.013$  & $0.02459 \pm 0.00085$ & 1.90 $^{+  1.41}_{-  1.22}$ &   0.81 $^{+  0.09}_{-  0.09}$\\[4pt]
MPG\,2.2\,m/GROND\,\#2 & Sloan $i^{\prime}$ & $0.1845 \pm 0.0063$ & $0.1347 \pm 0.0017$ & $84.33 \pm 0.56$ & $0.1626 \pm 0.0053$ & $0.02190 \pm 0.00093$ & 1.19 $^{+  0.93}_{-  0.75}$ &   0.40 $^{+  0.05}_{-  0.06}$\\[4pt]
MPG\,2.2\,m/GROND\,\#2 & Sloan $z^{\prime}$ & $0.2031 \pm 0.0078$ & $0.1334 \pm 0.0014$ & $82.77 \pm 0.56$ & $0.1792 \pm 0.0067$ & $0.0239~  \pm 0.0011$  & 2.97 $^{+  1.62}_{-  1.45}$ &   0.73 $^{+  0.11}_{-  0.12}$\\[6pt]
MPG\,2.2\,m/GROND\,\#3 & Sloan $g^{\prime}$ & $0.194  ~\pm 0.019$  & $0.1361 \pm 0.0038$ & $83.29 \pm 1.22$ & $0.171~  \pm 0.016$  & $0.0233~  \pm 0.0028$  & 4.59 $^{+  0.67}_{-  0.83}$ &   0.22 $^{+  0.16}_{-  0.15}$\\[4pt]
MPG\,2.2\,m/GROND\,\#3 & Sloan $r^{\prime}$ & $0.2076 \pm 0.0057$ & $0.1410 \pm 0.0014$ & $82.52 \pm 0.40$ & $0.1819 \pm 0.0048$ & $0.02567 \pm 0.00084$ & 1.34 $^{+  1.33}_{-  0.92}$ &   0.72 $^{+  0.08}_{-  0.07}$\\[4pt]
MPG\,2.2\,m/GROND\,\#3 & Sloan $i^{\prime}$ & $0.214 ~ \pm 0.014$  & $0.1406 \pm 0.0021$ & $82.13 \pm 0.88$ & $0.188~  \pm 0.012$  & $0.0264~  \pm 0.0020$  & 3.45 $^{+  0.46}_{-  0.53}$ &   0.14 $^{+  0.12}_{-  0.09}$\\[4pt]
MPG\,2.2\,m/GROND\,\#3 & Sloan $z^{\prime}$ & $0.1958 \pm 0.0084$ & $0.1376 \pm 0.0016$ & $83.19 \pm 0.61$ & $0.1721 \pm 0.0072$ & $0.0237~  \pm 0.0012$  & 3.59 $^{+  1.24}_{-  1.10}$ &   0.45 $^{+  0.11}_{-  0.14}$\\[6pt]
MPG\,2.2\,m/GROND\,\#4 & Sloan $g^{\prime}$ & $0.1958 \pm 0.0057$ & $0.144 ~ \pm 0.011$  & $82.88 \pm 0.33$ & $0.1711 \pm 0.0059$ & $0.0247~  \pm 0.0012$  & 3.15 $^{+  1.28}_{-  1.12}$ &   0.42 $^{+  0.10}_{-  0.15}$\\[4pt]
MPG\,2.2\,m/GROND\,\#4 & Sloan $r^{\prime}$ & $0.1962 \pm 0.0061$ & $0.1357 \pm 0.0074$ & $83.21 \pm 0.42$ & $0.1727 \pm 0.0062$ & $0.02344 \pm 0.00083$ & 3.09 $^{+  1.80}_{-  1.60}$ &   0.54 $^{+  0.11}_{-  0.18}$\\[4pt]
MPG\,2.2\,m/GROND\,\#4 & Sloan $i^{\prime}$ & $0.194  ~\pm 0.011$  & $0.138 ~ \pm 0.011$  & $83.21 \pm 0.71$ & $0.1704 \pm 0.0097$ & $0.0234~  \pm 0.0015$  & 2.00 $^{+  0.80}_{-  0.66}$ &   0.27 $^{+  0.06}_{-  0.09}$\\[4pt]
MPG\,2.2\,m/GROND\,\#4 & Sloan $z^{\prime}$ & $0.1955 \pm 0.0078$ & $0.1414 \pm 0.0089$ & $83.02 \pm 0.47$ & $0.1713 \pm 0.0076$ & $0.0242~  \pm 0.0012$  & 3.61 $^{+  1.26}_{-  1.15}$ &   0.46 $^{+  0.11}_{-  0.15}$\\[6pt]
CA 1.23\,m             & Cousins\,$R$       & $0.192 ~ \pm 0.0073$ & $0.1344 \pm 0.0012$ & $83.41 \pm 0.62$ & $0.1692 \pm 0.0062$ & $0.02275 \pm 0.00099$ & 4.27 $^{+  1.80}_{-  1.91}$ &   0.80 $^{+  0.09}_{-  0.10}$ \\[4pt]
Euler 1.2\,m           & Gunn\,$r$          & $0.1902 \pm 0.0072$ & $0.1397 \pm 0.0022$ & $83.77 \pm 0.63$ & $0.1669 \pm 0.0060$ & $0.023~3  \pm 0.0012$  & 2.23 $^{+  1.79}_{-  1.47}$ &   0.86 $^{+  0.07}_{-  0.07}$\\
\hline
{\bf Final results} & & $\mathbf{0.1947 \pm 0.0018}$ & $\mathbf{0.13677 \pm 0.00056}$ & $\mathbf{83.15 \pm 0.13}$ & $\mathbf{0.1710 \pm 0.0016}$ & $\mathbf{0.02369 \pm 0.00027}$ \\
\hline
\citet{smith:2012} & & ~~~~~~~~$-$ & $0.13842 \pm 0.00072$ & $83.61 \pm 0.21$ & ~~~~~~~~$-$ & ~~~~~~~~$-$ & &  \\[2pt]
\citet{maciejewski:2016} & & ~~~~~~~~$-$ & $0.1391_{-0.0012}^{+0.0011} $ & $83.62_{-0.26}^{+0.30} $ & ~~~~~~~~$-$ & ~~~~~~~~$-$ & & \\
\hline \end{tabular}
}\end{table*}

\section{Physical properties}
\label{sec_4}

We have determined the physical properties of the system using the {\it Homogeneous Studies} approach \citep{southworth:2012a}. The input quantities to this analysis were the photometric parameters (Sect.\,\ref{sec_3.2}) and the spectroscopic properties of the host star taken from \citet{smith:2012} (effective temperature $T_{\mathrm{eff}}=5959 \pm 134$\,K, metallicity $\FeH = -0.26 \pm 0.10$ and velocity amplitude $K_{\mathrm{A}}=391.5 \pm 8.3$\,m\,s$^{-1}$). These were augmented by an estimate of the velocity amplitude of the planet, $K_{\rm b}$, and the physical properties of the system calculated.

$K_{\rm b}$ was iteratively adjusted to maximise the agreement between the measured $R_{\mathrm{A}}/a$ and $T_{\mathrm{eff}}$ and those predicted by a set of theoretical models, considering a wide range of possible ages for the host star. Five sets of theoretical models were used (Claret: \citealp{claret:2004}; Y2: \citealp{Demarque:2004}; BaSTI: \citealp{pietrinferni:2004}; VRSS: \citealp{vandenberg:2006}; DSEP: \citealp{dotter:2008}), yielding five different estimates of each output quantity. We took the unweighted mean of these as the final values, and assigned a systematic error based on the level of agreement among the values obtained using different theoretical models. Statistical errors were propagated through the analysis from the input parameters. The final values are given in Table\,\ref{tab:finalparameters}. Our results are in good agreement with those found by \citet{smith:2012} and \citet{maciejewski:2016}, but are more precise and robust.
\begin{table*} \centering
\caption{Physical parameters of the planetary system WASP-36 derived in this work and compared with results from the literature. Where two errorbars
are given, the first refers to the statistical uncertainties and the second to the systematic errors.}
\label{tab:finalparameters}
\begin{tabular}{l c c c c c} \hline
Quantity & Symbol & Unit & This work & \citet{smith:2012} & \citet{maciejewski:2016}\\
\hline
Stellar mass                  & $M_{\rm A}$     & \Msun & $1.081 \pm 0.025 \pm 0.023$       & $1.040 \pm 0.031$   & $-$  \\
Stellar radius                & $R_{\rm A}$     & \Rsun & $0.985 \pm 0.012 \pm 0.007$       & $0.951 \pm 0.018$  & $0.960_{-0.019}^{+0.020}$   \\ [2pt]
Stellar surface gravity       & $\log g_{\rm A}$& cgs   & $4.486 \pm 0.009 \pm 0.003$       & $4.499 \pm 0.012$  &  $4.490_{-0.024}^{+0.026}$  \\[ 2pt]
Stellar density               & $\rho_{\rm A}$  & \psun & $1.132 \pm 0.032$                 & $1.211 \pm 0.050$  & $1.176_{-0.066}^{+0.060}$    \\ [4pt]
Planetary mass                & $M_{\rm b}$     & \Mjup & $2.361 \pm 0.062 \pm 0.033$       & $2.303 \pm 0.068$  & $2.295 \pm 0.058$  \\
Planetary radius              & $R_{\rm b}$     & \Rjup & $1.327 \pm 0.019 \pm 0.009$       & $1.281 \pm 0.029$  & $1.330_{-0.029}^{+0.030}$   \\[2pt]
Planetary surface gravity     & $g_{\rm b}$     & \mss  & $33.2 \pm  1.1$                   & $32.1 \pm  1.3$  & $33.7_{-1.4}^{+1.5}$     \\ [2pt]
Planetary density             & $\rho_{\rm b}$  & \pjup & $0.945 \pm 0.041 \pm 0.007$       & $1.096 \pm 0.067$  & $0.976_{-0.068}^{+0.070}$      \\[2pt]
Equilibrium temperature       & \Teq            & K     & $1733 \pm 19$                     & $1724 \pm 43$    & $-$     \\
Safronov number               & \safronov\      &       & $0.0880 \pm 0.0023 \pm 0.0006$    & $-$  & $-$                    \\
Orbital semi-major axis       & $a$             & au    & $0.02677 \pm 0.00021 \pm 0.00019$ & $0.02643 \pm 0.00026$ & $0.02641 \pm 0.00026$ \\
Age                           & $\tau$          & Gyr   & $1.4_{-0.3\,-3.4}^{+0.4\,+1.4}$   & $2.5^{+3.5}_{-2.2}$ & $-$  \\
\hline \end{tabular} \end{table*}

\section{Variation of the planetary radius with wavelength}
\label{sec_5}

Being strongly irradiated by their parent stars, the spectra of hot Jupiters are expected to show characteristic absorption features at optical wavelengths. Some of the predicted features include sodium ($\sim 590$\,nm), potassium ($\sim 770$\,nm), water vapour ($\sim 950$\,nm), and Rayleigh scattering at bluer wavelengths. However, the variety of hot-Jupiter transmission spectra suggests a great deal of variation in chemistry and atmospheric dynamics. While \citet{fortney:2005} suggested that a physical dichotomy exists between, essentially, insulated and non-insulated and well-mixed atmospheres (based primarily upon the level of irradiation received by the planet) the species of clouds driving the insulation has been disputed (e.g.\ \citealp{fortney:2008,zahnle:2009,knutson:2010}).

\begin{figure} \centering
\includegraphics[width=8.8cm]{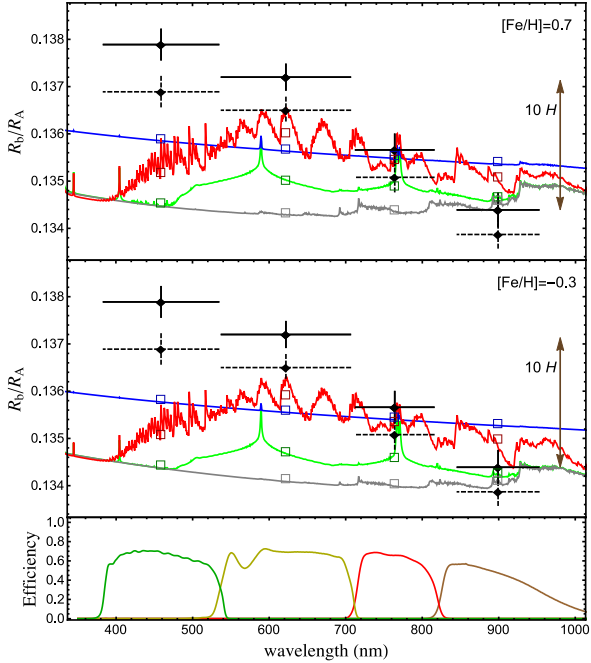}
\caption{Variation of the ratio of the radii with wavelength. The black points are the weighted mean $k$ values
from the GROND observations. The vertical bars represent the relative uncertainties and the horizontal bars show
the FWHM transmission of the passbands. The black points with dashed errorbars refer to the same measurements
after correction for possible starspot activity. Synthetic spectra for WASP-36\,b, obtained with the \emph{petitCODE},
are shown as coloured lines. The green line is a standard prediction, the grey line shows the case in which Na and K
are excluded, the blue line is as the grey one but with Rayleigh scattering increased by a factor of 1000, and the
red line shows the case in which we consider the opacity contribution due to TiO and VO. Offsets are applied to the
models to provide the best fit to our radius measurements. The atmospheres were computed for two different planetary
metallicities: [Fe/H]$_{\rm b} = -0.3$ ({\it top panel}) and [Fe/H]$_{\rm b} = 0.7$ ({\it bottom panel}). Coloured boxes
represent the predicted values for the three models integrated over the passbands of the observations. Transmission
curves of the GROND filters are shown in the bottom panel. The size of ten atmospheric pressure scale heights
(10\,$H$) is shown on the right of the plot.}
\label{fig:radius_variation}
\end{figure}

The ability of GROND to observe simultaneously in four optical bands makes it a useful tool to detect absorption features and thus probe the atmospheric composition of the planet, by measuring the ratio of the radii in different bands and comparing these values with synthetic spectra. Following the approach of \citet{southworth:2015}, we calculated the ratio of the radii in each passband with the other photometric parameters fixed to the best-fit values. This yielded a set of $k$ values which are directly comparable and whose errorbars exclude common sources of uncertainty. The errorbars were calculated using 10\,000 Monte Carlo simulations. We found $k_{g^{\prime}}=0.13789\pm0.00032$, $k_{r^{\prime}}= 0.13720 \pm 0.00027$, $k_{i^{\prime}}= 0.13566 \pm 0.00033$ and $k_{z^{\prime}}=0.13439 \pm 0.00035$ (Fig.\,\ref{fig:radius_variation}).

A significant variation of the planetary radius was found between the $g^{\prime}$ and the $z^{\prime}$ bands. The variation is roughly $11$ pressure scale heights\footnote{The pressure scale height is defined as $H = \frac{k_{\rm B}T^{\prime}_{\rm eq}}{\mu_{\mathrm{m}}\,g_{\mathrm{p}}}$, where $k_{\rm B}$ is Boltzmann's constant and $\mu_{\mathrm{m}}$ the mean molecular weight.} and is significant at beyond the 5$\sigma$ level. A similar variation has been found in other cases (e.g.\ HD\,189733\,b, \citealp{sing:2011}; HAT-P-5\,b, \citealp{southworth:2012b}; GJ\,3470\,b, \citealp{nascimbeni:2013}; Qatar-2\,b, \citealp{mancini:2014c}) and could be a sign of Rayleigh-scattering processes in the planetary atmosphere.

\begin{figure*} \centering
\includegraphics[width=\textwidth]{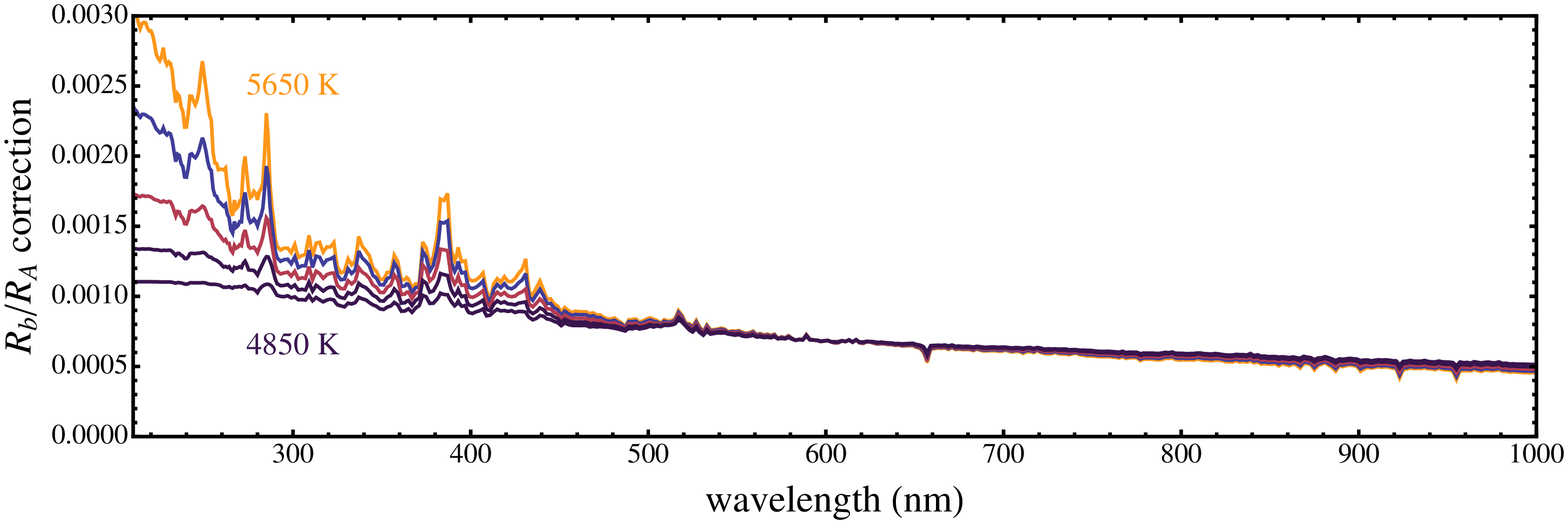}
\caption{The effect of unocculted starspots on the transmission spectrum of WASP-36\,b, considering a 1\% flux
drop at 600\,nm. We adopted a stellar temperature of $T_{\rm eff} = 5960$\,K, while the starspot coverage was
modelled using a grid of stellar atmospheric models of different temperature ranging from 5650\,K (yellow line)
to 4850\,K (purple line), in steps of 250\,K.}
\label{fig:correction}
\end{figure*}

We also considered the possibility that unocculted starspots, which have a stronger effect at bluer wavelengths, can cause a larger radius of WASP-36\,b in the blue band than that predicted by theoretical models. We estimated the effect of unocculted starspots on the transmission spectrum of WASP-36\,b using the methodology described by \citet{sing:2011}. The correction to the transit depth for unocculted spots is plotted in Fig.\,\ref{fig:correction}, assuming a total dimming of 1\% at a reference wavelength of 600\,nm \citep{sing:2011} for different starspot temperatures. Such corrections have been applied to the values of $k$ reported in Fig.\,\ref{fig:radius_variation} for each of the four optical band, but the difference between the blue and the red bands remains substantial.

In order to investigate this variation, we have compared the observed transmission spectrum of WASP-36\,b with sets of synthetic transmission spectra. These have been obtained by using two different codes, \emph{petitCODE} and {\sc vstar}. These two codes and the various models are described in the next two subsections.

\subsection{petitCODE}
\label{petitcode}

For calculating synthetic transmission spectra we used \emph{petitCODE} for self-consistent modelling of one-dimensional atmospheric structures and spectra \citep{mollierevanboekel2015}. \emph{petitCODE} has recently been extended for deriving transmission spectra by directly calculating the transmission through planetary annuli as probed during transits. An effective planetary radius is then calculated from the combined transmission of all annuli. The transmission spectra include Rayleigh scattering of H$_2$ and He, using the cross-sections reported in \citet{dalgarnowilliams1962} and \citet{chandalgarno:1965}, respectively. For verification purposes of our implementation of the transmission-spectrum calculations, we compared to the one-dimensional transmission spectra given in figs.\ 2 and 3 in \citet{fortney:2010}, which yielded excellent agreement. We further added molecular cross-sections for TiO and VO, using an updated line list based on \citet{plez:1998}, available on the author's website\footnote{\url{http://www.pages-perso-bertrand-plez.univ-montp2.fr/}} for TiO and a line list obtained from Betrand Plez (private communication) for VO. The partition functions were obtained from Uffe Gr\r{a}e J{\o}rgensen's website\footnote{\url{http://www.astro.ku.dk/~uffegj/scan/scan_tio.pdf}} for TiO and Betrand Plez (private communication) for VO, which is based on an updated partition function by \citet{sauvaltatum:1984}; see also \citet{gustafssonedvardsson:2008}. As pressure broadening information was not available, we approximated the broadening coefficients using Equation (15) in \citet{sharpburrows2007}. For this set of models we use an extended list of condensable species, comprising MgSiO$_3$, Mg$_2$SiO$_4$, SiC, Fe, Al$_2$O$_3$, Na$_2$S, KCl, H$_2$O, TiO and VO. The chemical equilibrium abundances for the gas and condensed phase species in \emph{petitCODE} are now calculated by a new chemical equilibrium code, which reliably works for temperatures between $60-20000$~K and has been tested for consistency with the {\sc cea} code \citep{gordon1994,mcbride1996}. For the development of this code we made use of the methods and equations outlined in the {\sc cea} manual \citep{mcbride1996}.

Using the values reported in Table\,\ref{tab:finalparameters} for the parameters of WASP-36\,b and its host star, we calculated self-consistent atmospheric structures. 
The stellar irradiation was calculated assuming a global average. As the flux is received by the planet with a cross-section of $\pi R_{\rm b}^2$, but is distributed over an area of $4\pi R_{\rm b}^2$, this corresponds to a flux dilution factor of $1/4$ when compared to the flux at the substellar point. We computed atmospheres for two different planetary metallicities: {[Fe/H]$_{\rm b}=-0.3$}, which is close to the value reported for the host star ($-0.26 \pm 0.10$, \citealt{smith:2012}), and [Fe/H]$_{\rm b}=0.7$, which corresponds to the planet being ten times more enriched than its host star. For both metallicities we calculated two atmospheres: in the first case we did not consider the opacities of TiO and VO in the atmospheric structure calculation, whereas in the second case we did consider such opacity contributions. For the cases in which TiO and VO were not considered, we took the resulting two pressure-temperature structures and calculated three different transmission spectra to investigate the following three cases:
\begin{itemize}
\item{nominal case, using the same opacities as were used for obtaining the atmospheric structure;}
\item{case without Na and K opacity;}
\item{case without Na and K opacity and a H$_2$ Rayleigh cross-section enhanced by a factor 1000 in order to mimic strong Rayleigh-like scatterers.}
\end{itemize}
In total we considered eight cases: $2\times 3$ cases for the atmospheric structures calculated without TiO and VO, plus two cases for the atmospheric structures including TiO and VO. They are depicted in Fig.\,\ref{fig:radius_variation}, and it is clear that none of the options explored above was able to reproduce the data, especially their steep slope.

The slope of the transit spectrum is given by \citep[see, e.g.,][]{etangspont2008}
\begin{equation}
\frac{dR_{\rm b}}{d\log(\lambda)}=\alpha H \,
\end{equation}
where $\alpha$ is the power law dependency of the opacity with respect to wavelength, $\alpha = d{\rm log}(\kappa) / d{\rm log}(\lambda)$, and $H$ is the atmospheric scale height. The surface gravity of WASP-36\,b is well constrained by its mass and radius measurements, and $\mu_{\rm m}$ should have a value close to 2.3\,amu if one assumes an atmosphere dominated by H$_2$ and He \citep{dewit:2013}. For WASP-36\,b the slope obtained for the synthetic transmission spectra is too small by a factor of $\sim 4$ when compared to the data. The minimum temperature in our atmospheric structure is $\sim 1000$\,K. Rayleigh scattering ($\alpha=-4$) would thus require a very high atmospheric temperature, of at least 4000 K, in order to be consistent with the data.
%

\subsection{VSTAR}
\label{vstar}

We created a second set of models of the atmosphere of WASP-36\,b with the Versatile Software for the Transfer of Atmospheric Radiation ({\sc vstar}; \citealp{bailey:2011}). {\sc vstar} is a robust line-by-line radiative transfer solver, which has been successfully applied to the atmospheres of cool stars, solar system planets, and hot Jupiters \citep{zhou:2013}. 
For WASP 36\,b, the Rayleigh and Mie scattering capabilities were used for attempting to model the dramatic increase with radius at bluer wavelengths. Rayleigh scattering from a H-He atmosphere alone is not able to describe the change in radius for WASP 36\,b that we have measured. A Mie scattering haze of small particles can be introduced to compensate for a large portion of the short wavelength absorption.

Such a scattering haze is introduced within the radiative transfer model and is dependent upon the optical depth per horizontal layer, the refractive indices of the scattering material across relevant wavelengths, and the radii of the particles themselves. Scattering atmospheric particles with radii ranging from 0.01 to 0.1\,$\mu$m were tested.  The distribution in sizes for a given median radius is described by a power law distribution with an effective variance of $0.01 \mu$m \citep{mishchenko:2002}.  Approximately half the slope can be derived with particle sizes around $0.03-0.04\,\mu$m. Changing the distribution of these particles vertically throughout the atmosphere can shape the Mie scattering contribution. In systems with sufficient data, this can be fit to provide information about the cloud and haze distribution on the planet.  The opaqueness of the layers set by the optical depth per height effects the spectrum particularly for upper layers where the transmission path length is longer.  Too great an optical depth at a given layer will create a flattened spectrum towards longer wavelengths masking molecular features.  Because of the shift in the relative radius from blue to red wavelengths for WASP 36b it is  unlikely that a very opaque cloud is present high in the atmosphere.

\begin{figure} \centering
\includegraphics[width=8.8cm]{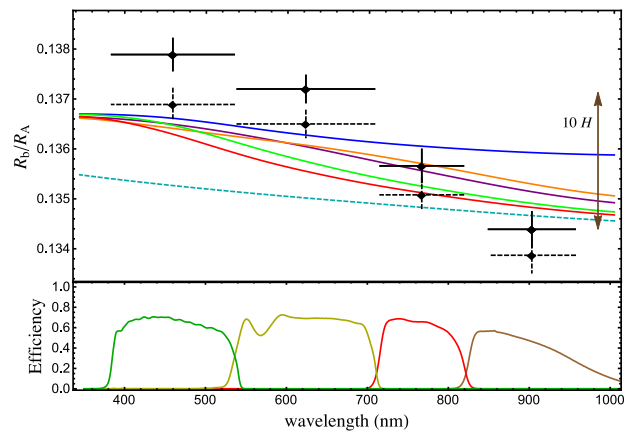}
\caption{Variation of the ratio of the radii with wavelength, same as in Fig.\,\ref{fig:radius_variation}.
Synthetic spectra for WASP-36\,b, obtained with the {\sc vstar} code (see Sect.\,\ref{vstar}), are shown
as coloured lines. The different lines correspond to varying optical depth with height, indicative of
the Mie scattering particle distribution. Pure Rayleigh scattering without the effects of clouds is
shown with the dashed cyan curve. No molecular absorption is included in these models.}
\label{fig:radius_variation2}
\end{figure}

In Fig.\,\ref{fig:radius_variation2} transmission spectral curves without molecular features are shown.  The variation in the radius is due solely to scattering by molecular species or clouds and hazes. 
The various lines correspond to different values of the optical depth in the upper atmosphere over 1000\,km. 
The model with H-He Rayleigh scattering alone (dashed cyan line) does not compensate for the steep rise in absorption towards blue wavelengths.  
The optical depth distributions shown are illustrative of the effects of clouds and hazes.  Blue scattering haze in a highly extended atmosphere could possibly produce the large scale height distribution of radii.

Layer-by-layer calculations for varying optical depth were trialled, but we found that they cannot compensate for the unusually steep absorption.  The scattering properties were based upon those of enstatite, which is a species likely responsible for the blue light scattering in the hot Jupiter HD\,189733\,b \citep{zahnle:2009}.  Other species with weaker attenuation at blue wavelengths, such as some sulphurous molecular species, may better compensate for the blue light absorption in the planet's transit. It is possible that along with strong Mie and Rayleigh scattering -- perhaps from an extended scattering haze -- that absorption in blue light from a molecular species is also present.

Relatively few species absorb in blue visible light wavelengths. \citet{zahnle:2009} showed that sulphanyl (HS) has an absorption spectrum which is effective at  short wavelengths, and they derived the absorption cross section for the species at some UV and blue wavelengths. Unfortunately no information on the species for longer wavelengths is currently available and we could not calculate the corresponding radiative transfer in the atmosphere of WASP-36\,b using the absorption cross section with {\sc vstar}.

\section{Conclusions}
\label{sec_6}

We have presented new broad-band photometric observations of five transit events in the WASP-36 planetary system, which is composed of a relatively young G2\,V star and a massive hot Jupiter. Four transits were observed with the GROND instrument, which supports observations in four optical bands simultaneously; another single-band transit observation was obtained with the CAHA 1.23\,m telescope. We fitted the light curves using the {\sc jktebop} code, and revised the orbital ephemeris and physical parameters of the system. Our results are consistent with and more precise than previous measurements.

We searched for a variation of the measured radius of WASP-36\,b as a function of wavelength, by determining the ratio of the planet's radius to that of the star in the four GROND passbands. Our data clearly show an increase of the planetary radius in the bluer bands. The variation between $g^{\prime}$ and $z^{\prime}$ is more than $10\,H$ with a significance level larger than 5$\sigma$.

We compared the multi-band measurements with the predictions of several theoretical models of planetary atmospheres. Synthetic spectra were computed using \emph{petitCODE} for self-consistent atmospheric structures and two opposite planetary metallicities; various cases with strong absorbers (TiO and VO), without Na and K opacity and with strong Rayleigh-like scattering process were investigated. Furthermore, the {\sc vstar} code was also utilised to produce atmospheric models in which the planetary-radius variation is only due to scattering by molecular species or clouds and hazes. Different models were presented varying the optical depth of the atmosphere. However, in no case was our models were able to reproduce the observed spectrum. Neither the presence of gaseous oxides nor strong Rayleigh scattering seems to be the cause of the steep slope in the transmission observations. The existence of an absorber which reproduces the observed transmission photometry due to its line opacities is therefore more likely, but the exact nature and origin of such an absorber is speculative.

In conclusion, the mechanism responsible for the steep slope in the transmission spectrum of WASP-36\,b remains difficult to constrain. Further observations of WASP-36\,b transits are suggested, especially with a photometer in the $U$ band or with a spectrograph able to cover a large spectral range at optical wavelengths.

\section*{Acknowledgements}

This paper is based on observations collected with the MPG 2.2\,m telescope located at the ESO Observatory in La Silla, Chile, and with the Zeiss 1.23\,m telescope at the Centro Astron\'{o}mico Hispano Alem\'{a}n (CAHA) at Calar Alto, Spain. Operation of the MPG 2.2\,m telescope is jointly performed by the Max Planck Gesellschaft and the European Southern Observatory. Operations at the Calar Alto telescopes are jointly performed by the Max Planck Institute for Astronomy (MPIA) and the Instituto de Astrof\'{i}sica de Andaluc\'{i}a (CSIC). GROND was built by the high-energy group of MPE in collaboration with the LSW Tautenburg and ESO, and is operated as a PI-instrument at the MPG 2.2\,m telescope. LM thanks Yan Betremieux for useful discussion. The reduced light curves presented in this work will be made available at the CDS (http://cdsweb.u-strasbg.fr/). We thank the anonymous referee for their useful criticisms and suggestions that helped us to improve the quality of this paper. The following internet-based resources were used in research for this paper: the ESO Digitized Sky Survey; the NASA Astrophysics Data System; the SIMBAD data base operated at CDS, Strasbourg, France; and the arXiv scientific paper preprint service operated by Cornell University.



\bsp 
\label{lastpage}
\end{document}